\documentclass[letterpaper, 10 pt, conference]{ieeeconf}  % Comment this line out if you need a4paper

\IEEEoverridecommandlockouts                              % This command is only needed if 
\usepackage{amsmath} 
\usepackage{amsfonts}
\usepackage{mathrsfs}
\usepackage{graphicx}
\usepackage{sgame}
\usepackage{soul} % hilighting 
\usepackage{cite}
\usepackage{verbatim} % block comments

\newtheorem{theorem}{Theorem}[section]
\newtheorem{proposition}[theorem]{Proposition}
\newtheorem{example}[theorem]{Example}

\newtheorem{lemma}[theorem]{Lemma}

\title{\LARGE\bf
On Partial Adoption of Vehicle-to-Vehicle Communication:\\ When Should Cars Warn Each Other of Hazards?
}

\author{Brendan T. Gould and Philip N. Brown% <-this % stops a space% 
\thanks{Brendan T. Gould and Philip N. Brown are with the department of Computer Science at the University of Colorado Colorado Springs, Colorado Springs, CO, USA {\tt\small \{bgould2,pbrown2\}@uccs.edu}}%
\thanks{This work was supported in part by the Undergraduate Research Academy at the University of Colorado Colorado Springs and in part by NSF Award \#ECCS-2013779.}
}

\DeclareMathOperator*{\argmin}{arg\,min}

\newcommand{\xn}{x_{\rm n}}
\newcommand{\xv}{x_{\rm v}}
\newcommand{\xvns}{x_{\rm vu}}
\newcommand{\xvs}{x_{\rm vs}}

\newcommand{\xne}{x^{\rm ne}}
\newcommand{\xnne}{x_{\rm n}^{\rm ne}}
\newcommand{\xvnsne}{x_{\rm vu}^{\rm ne}}
\newcommand{\xvsne}{x_{\rm vs}^{\rm ne}}

\newcommand{\cn}{\chi_{\rm n}}
\newcommand{\cvns}{\chi_{\rm vu}}

\newcommand{\Jn}{J_{\rm n}}
\newcommand{\Jvns}{J_{\rm vu}}
\newcommand{\Jvs}{J_{\rm vs}}

\newcommand{\A}{{\rm A}}
\renewcommand{\S}{{\rm S}} % overwrites the paragraph symbol

\newcommand{\C}{{\rm C}}
\newcommand{\R}{{\rm R}}

\begin{document}

\maketitle
\thispagestyle{empty}
\pagestyle{empty}

%%%%%%%%%%%%%%%%%%%%%%%%%%%%%%%%%%%%%%%%%%%%%%%%%%%%%%%%%%%%%%%%%%%%%%%%%%%%%%%%
\begin{abstract}
The emerging technology of Vehicle-to-Vehicle (V2V) communication over vehicular \emph{ad hoc} networks promises to improve road safety by allowing vehicles to autonomously warn each other of road hazards.
However, research on other transportation information systems has shown that informing only a subset of drivers of road conditions may have a perverse effect of increasing congestion.
In the context of a simple (yet novel) model of V2V hazard information sharing, we ask whether partial adoption of this technology can similarly lead to undesirable outcomes.
In our model, drivers individually choose how recklessly to behave as a function of information received from other V2V-enabled cars, and the resulting aggregate behavior influences the likelihood of accidents (and thus the information propagated by the vehicular network).
We fully characterize the game-theoretic equilibria of this model.
Our model indicates that for a wide range of our parameter space, V2V information sharing surprisingly increases the equilibrium frequency of accidents relative to no V2V information sharing, and that it may increase equilibrium social cost as well.
% in most cases, maximal hazard information sharing decreases both accident frequency and social cost.
% However, we identify parameter regimes with low V2V penetration where increasing the amount of information sharing among vehicles increases the probability of accidents or increases the social cost at equilibrium.
% Additionally, for a much larger portion of our parameter space, we show that the frequency of accidents increases with V2V mass, indicating that V2V technology tends to make drivers more reckless. 
\end{abstract}

\section{Introduction}
\label{sec:intro}
Technology is becoming increasingly intertwined with the society it serves, accelerated by emerging paradigms such as the internet of things (IoT) and various smart infrastructure concepts such as vehicle-to-vehicle communication (V2V).
It is no longer appropriate to design the merely-technical aspects of systems in isolation; rather, engineers must explicitly consider the implicit feedback loop between designed autonomy and human decision-making.
As a piece of this process, recent research has asked when new technological solutions may cause more harm than good~\cite{Brown2020a}.

A clear example of this is the area of equilibrium traffic congestion under selfish individual behavior. 
This topic has been well researched, and it is commonly understood that equilibria associated with this behavior may not be optimal at the system level~\cite{gairing_selfish_2008, correa_selfish_2004, dafermos_traffic_1984, wardrop_road_1952, massicot_public_2019, wu_information_2019}.
Many proposed solutions to this problem focus on the effects of deploying smart infrastructure to alleviate congestion and safety issues, using incentive design~\cite{Ferguson2021, lazar_learning_2021, biyik_incentivizing_2021} and information design~\cite{Zhu2020,massicot_competitive_2021} to improve upon selfish network routing.
However, this technology does not always have its intended effect; for example, self-driving cars can exacerbate equilibrium traffic congestion~\cite{Mehr2019}.

We study a common paradox of information design. 
\emph{Bayesian persuasion} describes the process of a sender disclosing or obfuscating information in an attempt to change the actions of other strategic agents~\cite{kamenica_bayesian_2011, bergemann_information_2019}.
% However, it is often difficult to anticipate the reactions of these agents to this information, giving rise to counter-intuitive results. 
However, it is known that merely making a subgroup of a population aware of a new road in a network can increase the equilibrium cost to that group, known as informational Braess' paradox~\cite{acemoglu_informational_2018, roman_how_2019}.
In information design problems in general, full disclosure of information is not always optimal~\cite{liu_effects_2016, massicot_public_2019, sayin_hierarchical_2019, tavafoghi_informational_2017,ukkusuri_information_2013,bergemann_information_2019}.
% These types of paradoxes are common throughout information sharing games; in general, 

This naturally gives rise to the question of ``What is the optimal information sharing policy?''
Prior research has posed this question in the context of congestion games where each driver's cost depends on the selected route and the total mass of drivers on that route \cite{ dafermos_traffic_1984, lazar_learning_2021, wu_information_2019, ben-akiva_dynamic_1991}.

In this paper, we initiate a study on the incentive effects of distributed hazard information sharing by V2V-equipped vehicles, and ask when maximal sharing optimizes driver safety.
We pose a simple model of information sharing with partial V2V adoption; that is, some vehicles are unable to receive signals warning of road hazards.
In contrast to existing literature, our model allows for \emph{endogenous} road hazards where the likelihood of a road hazard is dependent on the aggregate recklessness of drivers.

% Overviews of this topic can be found in
% A thorough discussion of the different forms this information could take can be found in 
% However, in contrast to the existing literature, our paper studies settings in which the probability of a road accident (and thus the probability that agents receive information regarding the accident) is endogenous; that is, dependent on the behavior of agents.
% Our contributions are unique in the following ways:
% first, we allow for an endogenous probability of an accident, determined by aggregate driver behavior. 
% Through a repeated cycle of many games, drivers can gradually adjust their behavior in response to the presence or absence of a warning signal indicating an accident.
% This creates a complex inter-relationship, which is captured by our concept of a signaling equilibrium.\looseness=-1

After fully characterizing the emergent behavior in terms of a new equilibrium concept we call a \emph{signaling} equilibrium (Theorem~\ref{thm:EQCharacterization}), our main result in Theorem~\ref{thm:signalingCrashProbOptimization} shows that there exist parameter regimes in which the optimal hazard signaling rate is 0; that is, sharing any road hazard information with only V2V-equipped vehicles leads to a higher frequency of accidents than sharing none.
We then close with a discussion of the relationship between the \emph{social cost} and the frequency of accidents, and show that these two objectives are sometimes fundamentally opposed to one another: a signaling policy which decreases the frequency of accidents may necessarily increase the social cost  (and vice-versa).\looseness=-1

\section{Model and Performance Metrics}
\label{section:model}

\subsection{Game Setup}
We adopt a nonatomic game formulation; i.e., we model a population of drivers as a continuum in which each of the infinitely-many drivers makes up an infinitesimally small portion of the population.
Each driver can choose to drive carefully (C), or recklessly (R), and a traffic accident either occurs ($\A$) or does not occur ($\neg \A$).
Reckless drivers become involved in existing accidents and experience an expected cost of $r>1$; however, careful drivers regret their caution if an accident is not present and experience a regret cost of $1$.
These costs are collected in this matrix:
\begin{center}
	\begin{game}{2}{2}
		\relax&Accident ($\A$)&No Accident ($\neg \A$)\\
		Careful ($\C$)&$0$&$1$\\
		Reckless ($\R$)&$r$&$0$\\
	\end{game}
\end{center}

We write $d$ to denote the total mass of drivers choosing to drive recklessly, and $p(d)$ to represent the resulting probability that an accident occurs. 
Throughout the manuscript, we assume that more reckless drivers make an accident strictly more likely, so that $p(d)$ is strictly increasing.

We model partial V2V penetration, i.e. some fraction $y~\in~[0, 1]$ of drivers have cars equipped with V2V technology.
If these drivers encounter a road hazard, V2V technology will autonomously detect this hazard and broadcast a warning signal with probability $q(y)$.
% These drivers may encounter road hazards and the technology may autonomously detect these hazards and broadcast warning signals. 
% We write $q(y)$ to denote the probability some V2V car detects an existing accident.
For simplicity, we assume that if a signal is broadcast, an accident must have occurred, and that the signal is received by every V2V car.\looseness=-1

% When a V2V car receives a signal, it may choose whether to display a warning light to its driver. 
In many models of transportation information systems, it is known that distributing perfect information can actually make parts or all of the population worse off~\cite{acemoglu_informational_2018, liu_effects_2016, sayin_hierarchical_2019, tavafoghi_informational_2017}.
Accordingly, we wish to study the information design problem faced by the administrators of V2V technology.
Therefore, let $\S$ be the event that a V2V car displays a warning to its driver, given that it has received a signal, and let $\mathbb{P}(\S) = \beta \in[0, 1]$.
Note that $\mathbb{P}(\A|\S) = 1$, since we have assumed there are no false positive warnings. 

This signaling scheme divides the population into three groups. 
We call a driver a \emph{non-V2V driver} if their vehicle lacks V2V technology, and a V2V driver otherwise.
We further differentiate V2V drivers by whether they have seen a warning signal, calling them \emph{unsignaled V2V drivers} and \emph{signaled V2V drivers}.
We write $\xn$, $\xvns$, and $\xvs$ to represent the mass of reckless drivers in each group, respectively, and a behavior profile as $x = (\xn, (\xvns, \xvs))$.

We model that both non-V2V and V2V drivers have habitual behaviors $\xn^*$ and $\xv^*$, respectively, and that these behaviors determine the probability of an accident. 
Every day, non-V2V cars follow this habitual behavior, i.e. $\xn = \xn^*$.
However, V2V drivers are able to adjust their behavior based on whether or not they see a warning on a particular day. 
The habitual behavior of V2V drivers must be a weighted average of their behavior when they do and do not see warnings, i.e.
\begin{equation}
    \xv^* = \mathbb{P}(\neg \S)\xvns + \mathbb{P}(\S)\xvs. \label{eq:avgBehaviorFull}
\end{equation}

In Lemma \ref{lemma:EQConditions}, we show that at equilibrium, $\xvs=0$. 
Therefore, in this manuscript we simplify \eqref{eq:avgBehaviorFull} to
\begin{equation}
    \xv^* = \mathbb{P}(\neg \S)\xvns. \label{eq:avgBehavior}
\end{equation}

Define $P(x)$ as the probability that an accident occurs, given some behavior profile $x$.
Additionally define $Q(x)$ as the probability that a specific V2V driver sees a warning light, given the same. 
When the dependence on $x$ is clear from context, we will sometimes write simply $P$ and $Q$. 
Then, $\mathbb{P}(\A) = P(x) = p(\xn^* + \xv^*)$ and $\mathbb{P}(\S) = Q(x) = P(x) \beta q(y)$.
Applying substitution with \eqref{eq:avgBehavior} gives that

\begin{equation}
    P(x) = p(\xn + (1-P(x)\beta q(y))\xvns). \label{eq:consistency}
\end{equation}

This uniquely specifies $P(x)$, since the left side is strictly increasing in $P$ and sweeps from $p(0)$ to $p(1)$, while the right side is strictly decreasing in $P$ and bounded within the same range.
% Henceforth, for the behavior profile $x$ we write $P(x)$ to denote the function implicitly defined by \eqref{eq:consistency}.
Note that $P(x)$ is implicitly parameterized by $\beta$ and $y$.\looseness=-1

We write $\Jn (a;x)$ to denote the expected cost to a non-V2V driver choosing action $a \in \{\C, \R\}$, and similarly $\Jvns (a;x)$ and $\Jvs (a;x)$ for unsignaled and signaled V2V drivers, respectively. 
These costs are given by
\begin{align}
    \Jn (a;x)&=\begin{cases}
        1-P(x) & \text{if $a=\C$},\\
        rP(x) & \text{if $a=\R$},
        \label{eq:NV2VAverageCost}
    \end{cases} \\
    \Jvns (a;x)&=\begin{cases}
        1 - \mathbb{P}(\A | \neg \S) & \text{if $a=\C$},\\
        r \mathbb{P}(\A | \neg \S) & \text{if $a=\R$},
        \label{eq:V2VUnsignaledCost}
    \end{cases} \\
    \Jvs (a;x)&=\begin{cases}
        0 & \text{if $a=\C$},\\
        r & \text{if $a=\R$}.
        \label{eq:V2VSignaledCost}
    \end{cases}
\end{align}
Note that \eqref{eq:V2VSignaledCost} holds due to our assumption that ${\mathbb{P} (\A | \S) = 1}$.
Finally, define a signaling game as the tuple $G~=~(\beta, y, r)$.

\subsection{Signaling Equilibrium}
We define a signaling equilibrium as a behavior profile $\xne~=~(\xnne, (\xvnsne, \xvsne))$ with $0 \le \xnne, \xvsne, \xvnsne \le 1$ satisfying the following:
\begin{align}
    \xnne < 1-y &\implies \Jn (\C; \xne) \leq \Jn (\R; \xne), \label{eq:NV2VCarefulInUse} \\
    \xnne > 0 &\implies \Jn (\R; \xne) \leq \Jn (\C; \xne), \label{eq:NV2VRecklessInUse} \\
    \xvnsne < y &\implies \Jvns (\C; \xne) \leq \Jvns (\R; \xne), \label{eq:V2VUnsignaledCarefulInUse} \\
    \xvnsne > 0 &\implies \Jvns (\R; \xne) \leq \Jvns (\C; \xne), \label{eq:V2VUnsignaledRecklessInUse} \\
    \xvsne < y &\implies \Jvs (\C; \xne) \leq \Jvs (\R; \xne), \label{eq:V2VSignaledCarefulInUse} \\
    \xvsne > 0 &\implies \Jvs (\R; \xne) \leq \Jvs (\C; \xne). \label{eq:V2VSignaledRecklessInUse}
\end{align}
Equations \eqref{eq:NV2VCarefulInUse}-\eqref{eq:V2VSignaledRecklessInUse} enforce the standard conditions of a Nash equilibrium (i.e. if players are choosing any action, its cost to them is minimal).
The novelty of this concept lies in the fact that we endogenously determine the mass of signaled and unsignaled V2V drivers using their behavior at equilibrium.

Additionally, we define social cost as the expected individual cost given behavior: 
\begin{equation}
\begin{split}
    S(x&) = \Jn (\C; x) (1 - y - \xn) + \Jn (\R; x) (\xn) + \\
    & (1 - Q) (\Jvns (\C; x) (y - \xvns) + \Jvns (\R; x) (\xvns)).
\end{split}
\label{eq:socialCost}
\end{equation}
Note that the decision of signaled V2V drivers is not included in this equation, since they will never incur a cost.

\begin{proposition}
For every signaling game $G$, there exists a signaling equilibrium $\xne$ and it is essentially unique.
By this we mean that for any two signaling equilibria $\xne_1$ and $\xne_2$ of $G$, both of the following hold:
\begin{align}
    \xne_{\rm{n}1} + (1-Q(\xne_1))\xne_{\rm{vu}1} &= \xne_{\rm{n}2} + (1-Q(\xne_2))\xne_{\rm{vu}2}, \label{eq:uniqueMass} \\
    P(\xne_1) &= P(\xne_2). \label{eq:uniqueCrashProb}
\end{align}
\end{proposition}
We provide a proof of this fact in Lemma \ref{lemma:EQTypes}.
Throughout this paper, we use the terms ``unique'' and ``essentially unique'' interchangeably to mean \eqref{eq:uniqueMass} and \eqref{eq:uniqueCrashProb} are satisfied. 

\subsection{Research Objectives}
Our first objective is to characterize the signaling equilibria of any game $G$.
In Theorem~\ref{thm:EQCharacterization}, we show that every game $G$ has an essentially unique signaling equilibrium. 
Additionally, we show that receiving a signal makes V2V drivers more cautious and not receiving a signal makes them more reckless at equilibrium, compared to non-V2V drivers.

Next, we seek to optimize accident probability and social cost by means of signal quality.
% Next, we introduce an optimization problem with respect to signal quality. 
% We consider optimization of accident probability and social cost, as they are both useful safety measures.
To that end, we abuse notation and write $P(G)$ to denote $P(\xne)$ and $S(G)$ to denote $S(\xne)$ where $\xne$ is a signaling equilibrium of game $G$.
We then wish to find values for $\beta_P$ and $\beta_S$ such that
\begin{gather}
    \beta_P \in \argmin_{\beta \in [0, 1]}{P(G)}, \label{eq:signalingCrashProbOptimization} \\
    \beta_S \in \argmin_{\beta \in [0, 1]}{S(G)}. \label{eq:signalingSocialCostOptimization}
\end{gather}

\begin{figure}
    \centering
    \vspace{2mm}
    \includegraphics[scale=0.6]{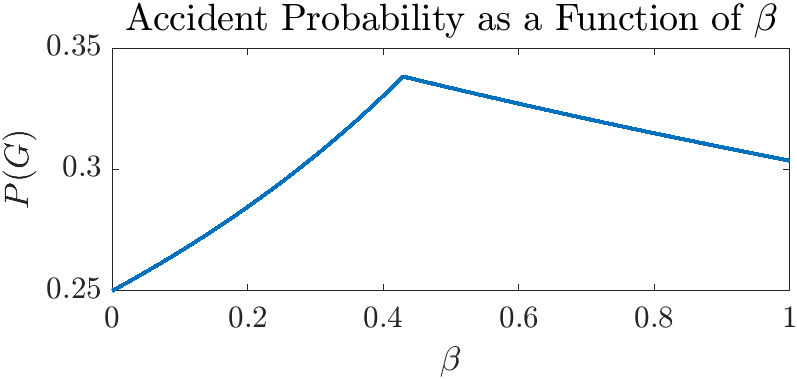}
    \caption{Equilibrium accident frequency with respect to the signal quality $\beta$.
    Note that when $\beta<0.43$, displaying more warning signals steeply increases the frequency of accidents, and that even the maximum possible signal quality $\beta=1$ yields a higher accident frequency than if warning signals were never shown to drivers.
    The example depicted has accident probability characterized by $p(d)=0.3d+0.1$, signal probability characterized by $q(y)=0.9y$, V2V penetration $y=0.9$, and accident cost $r=3$.
    \label{fig:PofG}}
    \vspace{-3mm}
\end{figure}

In Theorem~\ref{thm:signalingCrashProbOptimization}, we provide a constant time algorithm to determine a solution to \eqref{eq:signalingCrashProbOptimization}, and show that there exist games where $\beta_P = 0$ is a solution, as depicted in Figure~\ref{fig:PofG}.
Furthermore, in Proposition~\ref{prop:signalingSocialCostOptimization}, we provide sufficient criteria for when $\beta_S = 1$ is a solution to \eqref{eq:signalingSocialCostOptimization}, and show that there paradoxically exist regions of the parameter space where $\beta_S = 1$ is \emph{not} a solution to \eqref{eq:signalingSocialCostOptimization}.

\section{Our Contributions}
\subsection{Equilibrium Characterization}
A signaling equilibrium takes the form of a tuple listing the mass of reckless drivers in each of our three population groups.
These masses implicitly determine an equilibrium crash probability through \eqref{eq:consistency}.
Though this relationship is complicated, our first theorem shows that an equilibrium is uniquely determined by any given parameter combination. 

\begin{theorem}
\label{thm:EQCharacterization}
For any V2V signaling game $G=(\beta,y,r)$, a signaling equilibrium exists and is essentially unique.
In particular, the equilibrium $\xne = (\xnne, (\xvnsne, \xvsne))$  can take one of the following 3 forms:
\begin{itemize}
    \item (0, (0, 0)),
    \item (0, ($\cvns$, 0)), for some $\cvns \in [0, y]$,
    \item ($\cn$, (y, 0)), for some $\cn \in [0, 1-y]$.
\end{itemize}
\end{theorem}

This captures several important characteristics of signaling equilibria. 
Chief among these is the fact that a signaling equilibrium exists and is essentially unique for all games $G = (\beta, y, r)$.
% This one-to-one correspondence is very useful for optimization purposes. 
Additionally, note that signaled V2V drivers are certain that an accident has occurred, and will therefore never choose to be reckless at equilibrium. 
Finally, unsignaled V2V drivers have an extra measure of confidence that an accident has \emph{not} occurred, and are therefore more likely to drive recklessly than non-V2V drivers.

The proof proceeds as follows:
Lemma \ref{lemma:EQConditions} provides conditions describing all signaling equilibria. 
Next, we divide our parameter space into five regions that can be analyzed individually, and Lemma~\ref{lemma:crashProbRanges} derives the possible values of $P$ in each region.
% For each region, we derive the possible values of $P$ in Lemma \ref{lemma:crashProbRanges}.
Finally, Lemma~\ref{lemma:EQTypes} uses these results to show that exactly one signaling equilibrium exists for every parameter combination.
% These equilibria are of the form claimed in Theorem \ref{thm:EQCharacterization}, completing the proof.
We provide the full proofs for Theorem~\ref{thm:EQCharacterization} in Section~\ref{sec:proofs}.

\subsection{Information Design for Minimizing Accident Probability}
\label{subsec:signalingOptimization}
Though it seems intuitively obvious that increasing the quality of information given to drivers would allow them to make more informed decisions and arrive at less costly outcomes, prior research has shown that this is not always the case 
% Indeed, it is possible that giving drivers higher quality information can actually make them worse off at equilibrium than they were before 
\cite{acemoglu_informational_2018, roman_how_2019}. 
Because of this, it is a non-trivial question for V2V administrators to decide the optimal quality of information to distribute. 
This quality will be bound by technical limitations, but within that range, administrators can freely manipulate it by purposefully not displaying warnings to some drivers that receive signals. 
Paradoxically, we show that ignoring accidents in this way can decrease accident probability at equilibrium.\looseness=-1

\begin{theorem}
\label{thm:signalingCrashProbOptimization}
For any signaling game $G = (\beta, y, r)$, we must have that either:
\begin{equation}
    0 \in \argmin_{\beta \in [0, 1]}{P(G)} \text{ or } 1 \in \argmin_{\beta \in [0, 1]}{P(G)}. \label{eq:signalingMinCrashProb} \\
\end{equation}
Furthermore, there exist signaling games where $\beta=1$ does not minimize accident probability.
\end{theorem}

In other words, the minimum accident probability is guaranteed to be caused by never displaying warnings, or by displaying them as often as technologically possible.

The proof proceeds as follows:
first, we show that the probability of an accident is weakly increasing for low values of $\beta$, and weakly decreasing otherwise. 
Then, the smallest possible value of $\beta$ will always be a minimum within the increasing range, and the largest value of $\beta$ must be a minimum in the decreasing range. 
Therefore, one of the two must be a global minimum, which is the desired result. 

We will now provide a full definition of $P(G)$, the function optimized by \eqref{eq:signalingCrashProbOptimization}.
Note that in general, $P(G) = P(\xne)$, but with the following regions, we can be more specific:
\begin{gather} % I use gather and not align b/c there is no nice way to line these up
p\left(1-\frac{\beta q(y)}{1+r}y\right) < \frac{1}{1+r} \label{range:NRVR}
\end{gather}
% --------------------------------------------------------------------
\begin{gather}
p\left(\left(1-\frac{\beta q(y)}{1+r}\right)y\right) \leq \frac{1}{1+r} \leq p\left(1-\frac{\beta q(y)}{1+r}y\right) \label{range:NIVR}
\end{gather}
% --------------------------------------------------------------------
\begin{gather}
% \end{gather}
% \begin{gather}
p(0)~>~\frac{1}{1+r(1-\beta q(y))} \label{range:NCVC}
\end{gather}
% --------------------------------------------------------------------
\begin{gather}
p(0) \leq \frac{1}{1+r(1-\beta q(y))} \leq p \hspace{-0.5mm} \left( \hspace{-0.5mm} \left(1 - \frac{\beta q(y)}{1+r(1-\beta q(y))} \hspace{-0.5mm} \right) \hspace{-0.3mm} y \hspace{-0.3mm} \right) \label{range:NCVI}
\end{gather}
% --------------------------------------------------------------------
\begin{gather}
p\left(\left(1 - \frac{\beta q(y)}{1+r(1-\beta q(y))} \right)y\right) < \frac{1}{1+r(1-\beta q(y))} \notag \\ 
\text{ and } \frac{1}{1+r} < p\left(\left(1-\frac{\beta q(y)}{1+r}\right)y\right) \label{range:NCVR}
\end{gather}

For any signaling game $G = (\beta, y, r)$, the probability of an accident at its signaling equilibrium is given by
\begin{equation}
    P(G) = \begin{cases}
        p(0) & \text{if \eqref{range:NCVC},} \\
        \frac{1}{1+r(1 - \beta q(y))} & \text{if \eqref{range:NCVI},} \\
        P(\xne) & \text{if \eqref{range:NCVR},} \\
        \frac{1}{1+r} & \text{if \eqref{range:NIVR},} \\
        P(\xne) & \text{if \eqref{range:NRVR},}
    \end{cases}
    \label{eq:crashProbs}
\end{equation}
where $\xne$ is the signaling equilibrium satisfying Lemma \ref{lemma:EQTypes}.
A proof of this simplification can be found in Lemma \ref{lemma:crashProbRanges}.
Since there exists a signaling equilibrium for any game $G$, this function is defined for all parameter combinations.

% Next, we show that this function is monotonic over two ranges with respect to $\beta$. 
% Unfortunately, these conditions do not permit a closed form expression where $\beta q(y)$ is isolated. 

\begin{lemma}
\label{lemma:signalingCrashProbMonotonicity}
For sufficiently low values of $\beta$, $P((\beta, y, r))$ is weakly increasing; otherwise it is weakly decreasing. 
\end{lemma}

\begin{proof}
For any combination of the parameters $y$ and $r$, let $\beta_1, \beta_2 \in [0, 1]$ and $\beta_1 < \beta_2$. 
Let $G_1 = (\beta_1, y, r)$ and $G_2 = (\beta_2, y, r)$.
% Note that $G_1$ and $G_2$ will each satisfy one of $\eqref{range:NRVR}-\eqref{range:NCVR}$.
Our approach is a exhaustive comparison of crash probabilities within and between the cases of \eqref{eq:crashProbs}. 
If
\begin{equation*}
    \frac{1}{1+r(1-\beta_2 q(y))} \leq p\left(\left(1 - \frac{\beta_2 q(y)}{1+r(1-\beta_2 q(y))} \right)y\right), 
\end{equation*}
then $G_1$ and $G_2$ satisfy either \eqref{range:NCVC} or \eqref{range:NCVI}. 
In any case, algebraic manipulations on \eqref{eq:crashProbs} give that $P(G_1) < P(G_2)$.

Otherwise, we have that
\begin{equation*}
    \frac{1}{1+r(1-\beta_1 q(y))} > p\left(\left(1 - \frac{\beta_1 q(y)}{1+r(1-\beta_1 q(y))} \right)y\right).
\end{equation*}

Then, $G_1$ and $G_2$ each satisfy \eqref{range:NCVR}, \eqref{range:NIVR}, or \eqref{range:NRVR}.
We claim that $P$ is decreasing with respect to $\beta$. 
Simple algebra and Lemma \ref{lemma:crashProbRanges} suffices to obtain this result in every case except the following.
If $G_1$ and $G_2$ satisfy \eqref{range:NRVR}, then Lemma \ref{lemma:EQTypes} guarantees that $(1 - y, y, 0)$ is an equilibrium. 
Then, by \eqref{eq:consistency},
\begin{equation*}
    P(x) = p(1 - y + (1 - P(x)\beta q(y))y) = p(1 - P(x) \beta q(y) y).
\end{equation*}

Let $P_1$ and $P_2$ be the quantities that satisfy ${P_1 = p(1 - P_1 \beta_1 q(y)y)}$ and ${P_2 = p(1 - P_2 \beta_2 q(y)y)}$, respectively. 
By \eqref{eq:consistency} and \eqref{eq:crashProbs}, $P(G_1) = P_1$ and $P(G_2) = P_2$.
Assume by way of contradiction that $P_1 \leq P_2$. 
We use algebraic manipulations to work ``up'' one level of recursion, starting with the definition of $\beta_1$ and $\beta_2$.
This gives that 
\begin{equation*}
    1 - P_1 \beta_1 q(y) y > 1 - P_2 \beta_2 q(y) y.
\end{equation*}
Since $p(d)$ is increasing, it preserves the inequality, so
\begin{equation*}
    p(1 - P_1 \beta_1 q(y) y) > p(1 - P_2 \beta_2 q(y) y).
\end{equation*}
But then by definition of $P_1$ and $P_2$, we can substitute to obtain $P_1 > P_2$, contradicting our hypothesis. 
Therefore, we must have that $P(G_1) = P_1 > P_2 = P(G_2)$, the desired conclusion. 
If both games satisfy \eqref{range:NCVR}, a very similar technique can be used. 
Thus, $P$ is decreasing with $\beta$. 
\end{proof}

This result immediately gives a minimizing signal quality in each range. 
We use this result to prove Theorem \ref{thm:signalingCrashProbOptimization}.

\subsubsection*{Proof of Theorem~\ref{thm:signalingCrashProbOptimization}}
Immediately from Lemma \ref{lemma:signalingCrashProbMonotonicity}, we have that either the smallest or largest value of $\beta$ must minimize $P$.
Therefore, either $0 \in \argmin_{\beta \in [0, 1]}{P(G)}$, or $1 \in \argmin_{\beta \in [0, 1]}{P(G)}$, as desired.

It remains to show that there actually exist signaling games such that $\beta_P = 1$ does not minimize accident probability. 
To that end, let $p(d) = 0.8d + 0.1$, $q(y) = 0.9y$, $y = 0.7$, and $r = 20$.
Then $G_0 = (0, y, r)$ satisfies \eqref{range:NCVC}, and $G_1 = (1, y, r)$ satisfies \eqref{range:NCVI}. 
Therefore, by \eqref{eq:crashProbs}, $P(G_0) = p(0) = 0.1$, and $P(G_1) = \frac{1}{1+r(1-(1)q(y)))} \approx 0.1190$. 
Since $P(G_0) < P(G_1)$, $\beta_P = 1$ cannot be a solution.
\hfill\QED

\subsection{Information Design for Minimizing Social Cost}
\label{subsec:signalingOptimizationSocialCost}

It is also useful to consider how to minimize social cost at equilibrium. 
Again, intuition suggests that the social cost minimizing value of $\beta$ should be $1$.
We present the counter-intuitive result that there exist games where full information sharing among V2V drivers does not optimize social cost.

We provide an example to illustrate that $\beta_S$ need not be 1.\looseness=-1
\begin{example}
Let $p(d) = d^{0.25}$, $q(y) = 0.9y$, $y = 0.07$, and $r = 1.001$. 
Additionally, let $\beta_1 = 0.9$, $\beta_2 = 1$, $G_1 = (\beta_1, y, r)$, and $G_2 = (\beta_2, y, r)$.
Then, $G_1$ and $G_2$ both satisfy \eqref{range:NCVR}.
Numerical solvers give that $S(G_1) \approx 0.4889$, while $S(G_2) \approx 0.4890$.
Thus, increasing the quality of V2V information can increase the expected cost to drivers. 
\end{example}

\begin{proposition}
\label{prop:signalingSocialCostOptimization}
For any signaling game $G = (\beta, y, r)$, $\beta_S = 1$ is a solution to \eqref{eq:signalingSocialCostOptimization} unless $G$ satisfies \eqref{range:NCVR}.
\end{proposition}

\subsubsection*{Proof Sketch}
For any of our equilibrium ranges, applying Lemmas \ref{lemma:EQConditions} and \ref{lemma:EQTypes} gives an algebraic expression for $S(G)$.
In all ranges except \eqref{range:NCVR}, the expressions can be easily shown to be decreasing in $\beta$. 
Thus, $\beta_S = 1$ is a solution to \eqref{eq:signalingSocialCostOptimization}.
\hfill \QED

Note that if $G$ satisfies \eqref{range:NCVR}, $S(G)$ can be increasing with respect to $\beta$, but $P(G)$ is guaranteed to be decreasing by Lemma \ref{lemma:signalingCrashProbMonotonicity}.
Additionally, if $G$ satisfies \eqref{range:NCVI}, then $P(G)$ is increasing and $S(G)$ is decreasing. 
This implies that V2V administrators face an inherent trade-off in their optimization decision. 
To minimize accident probability, they must sometimes accept a higher than optimal social cost, and vice versa.\looseness=-1

\section{Proofs of Theorem~\ref{thm:EQCharacterization}} \label{sec:proofs}

\begin{lemma}
\label{lemma:EQConditions}
For any signaling game $G = (\beta, y, r)$, a behavior profile $\xne = (\xnne, (\xvnsne, \xvsne))$ is a signaling equilibrium if $\xvsne = 0$, and the following hold:
\begin{align}
    \xnne &= \begin{cases}
        0 & \text{if } P(\xne) > \frac{1}{1+r}, \\
        p^{-1}(P(\xne)) - (1 - Q(\xne))y & \text{if } P(\xne) = \frac{1}{1+r}, \\
        1-y & \text{if } P(\xne) < \frac{1}{1+r},
        \label{eq:NV2VChoice}
    \end{cases} \\
    \xvnsne &= \begin{cases}
        0 & \text{if } \mathbb{P} (\A | \neg \S)  > \frac{1}{1+r}, \\
        \frac{p^{-1}(P(\xne))}{1-Q(\xne)} & \text{if } \mathbb{P} (\A | \neg \S)  = \frac{1}{1+r}, \\
        y & \text{if } \mathbb{P} (\A | \neg \S)  < \frac{1}{1+r}.
        \label{eq:V2VUnsignaledChoice}
    \end{cases}
\end{align}

If $\beta q(y) > 0$, then this is the only signaling equilibrium. 
Otherwise, all signaling equilibria that exist satisfy \eqref{eq:uniqueMass} and \eqref{eq:uniqueCrashProb}. 
That is, this equilibrium is essentially unique. 
\end{lemma}

The proof of Lemma \ref{lemma:EQConditions} is simple, but notationally intensive; for brevity, we provide a proof sketch here.
A relationship which is useful in this section is:
\begin{equation}
    \label{eq:V2VConfidence}
    \frac{1}{1+r} \leq \frac{1}{1+r(1 - \beta q(y))}.
\end{equation}

Additionally, it can be shown using Bayes' Theorem that, 
\begin{equation}
    \def\arraystretch{0.5}
    \label{eq:V2VUnsignaledBayes}
    \mathbb{P} (\A | \neg \S ) 
    \begin{array}{c}
    <  \\ = \\ >
    \end{array}
    \frac{1}{1+r} \hspace{-0.8mm} \iff \hspace{-0.8mm} P(x)
    \begin{array}{c}
    < \\ = \\ >
    \end{array}
     \frac{1}{1+r(1-\beta q(y))}.
     \def\arraystretch{1.0}
\end{equation}

We use this notation to mean that any of the relationships between the first expressions is equivalent to the corresponding relationship between the second expressions. 
Equality and both inequalities are preserved. 

\subsubsection*{Proof Sketch}
In the forward direction, simple algebra largely suffices. 
For each case of \eqref{eq:NV2VChoice} and \eqref{eq:V2VUnsignaledChoice}, the necessary values of $P$ and $\mathbb{P}(\A|\neg\S)$ can be plugged into \eqref{eq:NV2VAverageCost} and \eqref{eq:V2VUnsignaledCost} to obtain a relationship on the expected costs to each group to drive carefully or recklessly.
This relationship will always satisfy \eqref{eq:NV2VCarefulInUse}-\eqref{eq:V2VUnsignaledRecklessInUse} for the corresponding value of $\xnne$ or $\xvnsne$.

Additionally, we always have that $\Jvs(\C) < \Jvs(\R)$, meaning that $\xvsne = 0$ will always satisfy \eqref{eq:V2VSignaledCarefulInUse} and \eqref{eq:V2VSignaledRecklessInUse}.
Therefore, the tuple described is a signaling equilibrium. 

The reverse direction is slightly more nuanced, but similar. 
The first and third cases of \eqref{eq:NV2VChoice} and \eqref{eq:V2VUnsignaledChoice}, as well as the condition that $\xvsne = 0$ are simple. 
The described values of $P$ and $\mathbb{P}(\A|\neg\S)$ imply a relationship between the cost of each behavior to non-V2V and unsignaled V2V drivers, respectively. 
After applying \eqref{eq:NV2VCarefulInUse}-\eqref{eq:V2VUnsignaledRecklessInUse}, this relationship contradicts all values of $\xnne$ and $\xvnsne$ except for those claimed in the Lemma. 
Similarly, \eqref{eq:V2VSignaledCost} immediately gives a relationship on costs that contradicts \eqref{eq:V2VSignaledCarefulInUse}-\eqref{eq:V2VSignaledRecklessInUse} for any values except $\xvsne = 0$.\looseness=-1

It remains to show the second cases of \eqref{eq:NV2VChoice} and \eqref{eq:V2VUnsignaledChoice}.
If $\beta q(y) > 0$, the the inequality described in \eqref{eq:V2VConfidence} becomes strict. 
Consider \eqref{eq:NV2VChoice} and assume $P = \frac{1}{1+r}$. 
Then, by \eqref{eq:V2VUnsignaledBayes} and the above, we know that $\xvnsne=y$.
By \eqref{eq:consistency}, this provides an algebraic expression for $P$ as a function of solely $\xnne$, which can be solved to produce the desired result. A proof of \eqref{eq:V2VUnsignaledChoice} is nearly identical. 

Otherwise, we must have that $\beta q(y) = 0$. 
In the second case of either \eqref{eq:NV2VChoice} or \eqref{eq:V2VUnsignaledChoice}, \eqref{eq:consistency} simplifies to
\begin{equation*}
    P(x) = p(\xnne + \xvnsne) = \frac{1}{1+r}.
\end{equation*}
This implies that all signaling equilibria that exist satisfy \eqref{eq:uniqueMass} and \eqref{eq:uniqueCrashProb}, so they are essentially identical.
Therefore, the tuple satisfying \eqref{eq:NV2VChoice} and \eqref{eq:V2VUnsignaledChoice} is a signaling equilibrium, and is essentially unique. 
\hfill \QED

Recall that \eqref{range:NRVR}-\eqref{range:NCVR} divide our parameter space into five regions. 
Lemma~\ref{lemma:crashProbRanges} shows that each of these regions restricts the possible values of $P$.

\begin{lemma}
\label{lemma:crashProbRanges}
For any signaling game $G = (\beta, y, r)$, at least one of \eqref{range:NRVR}-\eqref{range:NCVR} is true, and
\begin{align}
\eqref{range:NRVR} &\implies P < \frac{1}{1+r}, \label{eq:NRVRCrashProbCond} \\
% --------------------------------------------------------------------
\eqref{range:NIVR} &\implies P = \frac{1}{1+r}, \label{eq:NIVRCrashProbCond} \\
% --------------------------------------------------------------------
\eqref{range:NCVC} &\implies P = p(0), \label{eq:NCVCCrashProbCond}
\end{align}
\begin{align}
% --------------------------------------------------------------------
\eqref{range:NCVI} &\implies P = \frac{1}{1+r(1-\beta q(y))}, \label{eq:NCVICrashProbCond} \\
% --------------------------------------------------------------------
\eqref{range:NCVR} &\implies \frac{1}{1+r} < P < \frac{1}{1+r(1-\beta q(y))}. \label{eq:NCVRCrashProbCond}
\end{align}
\end{lemma}

The five claims are proved via contradiction. 
Applying Lemma \ref{lemma:EQConditions} to the contradiction hypothesis gives the only possible equilibrium tuple $\xne$.
Next, using this tuple and \eqref{eq:consistency}, we perform algebraic operations to take $P$ ``up'' one level of its recursive definition. 
Finally, we show that this new expression for $P$ forces a contradiction. 

\begin{proof}
First note that \eqref{range:NCVR} is simply the compliment of the other conditions, meaning that at least one must be true for any game $G$.

Since the remaining technique is identical across all five claims, we will prove only \eqref{eq:NIVRCrashProbCond} for brevity. 
Consider the case where $G$ is in the range defined by defined by \eqref{range:NIVR}, i.e.
\begin{equation*}
    p\left(\left(1-\frac{\beta q(y)}{1+r}\right)y\right) \leq \frac{1}{1+r} \leq p\left(1-\frac{\beta q(y)}{1+r}y\right).
\end{equation*}
If we assume by way of contradiction that $P(\xne) < \frac{1}{1+r}$, by \eqref{eq:V2VConfidence} and \eqref{eq:V2VUnsignaledBayes}, $\mathbb{P} (\A | \neg \S) < \frac{1}{1+r}$. 
Therefore, by Lemma \ref{lemma:EQConditions}, $\xnne = 1-y$, $\xvnsne = y$, and $\xvsne = 0$.
We then substitute into \eqref{eq:consistency} to obtain $P(\xne) = p(1 - y + (1 - P \beta q(y))y + 0) = p(1 - P \beta q(y)y)$.

Then, starting with our contradiction hypothesis, we perform algebraic operations to take $P$ ``up'' one level of this recursive definition. 
This gives that 
\begin{equation*}
    1 - \frac{\beta q(y)}{1+r}y < 1 - P \beta q(y)y.
\end{equation*}

Since $p(d)$ is strictly increasing, it preserves the inequality, giving
\begin{equation*}
    p\left(1 - \frac{\beta q(y)}{1+r}y\right) < p(1 - P \beta q(y)y).
\end{equation*}
Therefore, applying \eqref{range:NIVR}, we have that 
\begin{equation*}
    \frac{1}{1+r} \leq p\left(1 - \frac{\beta q(y)}{1+r}y\right) < p(1 - P \beta q(y)y) = P < \frac{1}{1+r},
\end{equation*}
an obvious contradiction. 
This technique can also be used to show that assuming $P > \frac{1}{1+r}$ forces a contradiction. 
Therefore, we must have that $P(\xne) = \frac{1}{1+r}$. 

A proof of the remaining claims can be accomplished in an identical manner.
\end{proof}

From Lemma \ref{lemma:crashProbRanges} we now know the possible long term accident probabilities in any region of parameter space. 
Based on these values, using Lemma \ref{lemma:EQConditions}, we can derive what a signaling equilibrium in each range must look like. 

\begin{lemma}
For any signaling game $G = (\beta, y, r)$, a unique signaling equilibrium $\xnne$ exists and takes the following form: 
\label{lemma:EQTypes}
\end{lemma}
\begin{align}
    \eqref{range:NRVR} &\implies \xne = (1-y, y, 0), \label{eq:NRVREqForm} \\
    % --------------------------------------------------------------------
    \eqref{range:NIVR}  &\implies \xne = \hspace{-0.8mm} \left( \hspace{-0.8mm} p^{-1} \hspace{-0.8mm} \left( \hspace{-0.8mm} \frac{1}{1+r} \hspace{-0.8mm} \right) \hspace{-0.8mm} - \hspace{-0.8mm} \left( \hspace{-0.8mm} 1 - \frac{\beta q(y)}{1+r} \hspace{-0.8mm} \right) \hspace{-0.8mm} y, y, 0 \hspace{-0.8mm} \right) \hspace{-0.8mm}, \label{eq:NIVREqForm} \\
    % --------------------------------------------------------------------
    \eqref{range:NCVC} &\implies \xne = (0, 0, 0), \label{eq:NCVCEqForm} 
    \end{align}
    \begin{align}
    % --------------------------------------------------------------------
    \eqref{range:NCVI} &\implies \xne = \left( 0, \frac{p^{-1}(\frac{1}{1+r(1-\beta q(y))})}{1-\frac{\beta q(y)}{1+r(1-\beta q(y))}}, 0 \right), \label{eq:NCVIEqForm} \\
    % --------------------------------------------------------------------
    \eqref{range:NCVR} &\implies \xne = (0, y, 0). \label{eq:NCVREqForm}
\end{align}

% For each of the five claims, we reuse the following proof method:
% \begin{enumerate}
%     % \item Assume the values of $\beta$, $y$, and $r$ are within the specific region
%     \item For each region, apply Lemma \ref{lemma:crashProbRanges} to obtain a condition on $P(x)$
%     \item Use \eqref{eq:V2VConfidence}, \eqref{eq:V2VUnsignaledBayes}, and the condition on $P(x)$ as needed to derive a similar condition on $\mathbb{P} (\A | \neg \S)$
%     \item Apply Lemma \ref{lemma:EQConditions} using these conditions to derive which type of equilibrium exists in that region
% \end{enumerate}

\begin{proof}
We prove this in cases. 
First, assume that $G$ is in the range defined by \eqref{range:NRVR} i.e.
\begin{equation*}
    p\left(1-\frac{\beta q(y)}{1+r}y\right) < \frac{1}{1+r}.
\end{equation*}
By Lemma \ref{lemma:crashProbRanges}, $P < \frac{1}{1+r}$.
Then, by \eqref{eq:V2VConfidence} and \eqref{eq:V2VUnsignaledBayes}, $\mathbb{P} (\A | \neg \S) < \frac{1}{1+r}$.
Finally, by Lemma \ref{lemma:EQConditions}, $(1 - y, y, 0)$ is a signaling equilibrium and essentially unique. 

An identical method can be used to show that $(0, 0, 0)$ and $(0, y, 0)$ are essentially unique signaling equilibria if $G$ is in the range of \eqref{range:NCVC} or \eqref{range:NCVR}, respectively. 

Now, assume that $G$ is in the range defined by \eqref{range:NIVR}.
By Lemma \ref{lemma:crashProbRanges}, $P(\xne) = \frac{1}{1+r}$. 
If $\beta q(y) > 0$, then $\frac{1}{1+r} < \frac{1}{1+r(1 - \beta q(y))}$, so by \eqref{eq:V2VUnsignaledBayes}, $\mathbb{P} (\A | \neg \S) < \frac{1}{1+r}$.
By Lemma \ref{lemma:EQConditions}, 
\begin{equation*}
    \left(p^{-1}\left(\frac{1}{1+r}\right) - \left(1 - \frac{\beta q(y)}{1+r}\right)y, y, 0\right)
\end{equation*}
is then an essentially unique signaling equilibrium. 
Otherwise, $\beta q(y) = 0$, so $Q(\xne) = 0$ and \eqref{eq:V2VUnsignaledBayes} implies that $\mathbb{P} (\A | \neg \S) = \frac{1}{1+r}$.
Therefore by Lemma \ref{lemma:EQConditions}, $\xnne = p^{-1}(P(\xne)) - (1 - Q(\xne))y = p^{-1}(\frac{1}{1+r}) - y$ and $\xvnsne = \frac{p^{-1}(P(\xne))}{1-Q(\xne)} = p^{-1} (\frac{1}{1+r})$. 
By \eqref{eq:consistency}, this gives that 
\begin{equation*}
    \frac{1}{1+r} = p \left(p^{-1} \left(\frac{1}{1+r} \right) - y + p^{-1} \left(\frac{1}{1+r} \right)\right),
\end{equation*}
forcing $p^{-1}(\frac{1}{1+r}) = y$.
Therefore, by substitution, $(0, y, 0)$ must be an essentially unique signaling equilibrium (note that this is a special case of the more general form given above). 
A similar technique can be used to show that \eqref{range:NCVI} forces a signaling equilibrium of the form claimed. 

By Lemma \ref{lemma:crashProbRanges}, any game $G$ must satisfy at least one of the above conditions, and therefore has an essentially unique signaling equilibrium. 
\end{proof}

Finally, we are equipped to prove Theorem \ref{thm:EQCharacterization}.

\subsubsection*{Proof of Theorem~\ref{thm:EQCharacterization}}
Lemma \ref{lemma:EQTypes} demonstrated existence and essential uniqueness of a signaling equilibrium for all signaling games $G$. 
Note that each of these equilibria are consistent with the forms claimed, completing the proof. 
\hfill\QED

\section{Conclusion}
This paper has posed and analyzed a simple model of self-interested driver behavior in the presence of road hazard signals.
We have shown that warning a subset of drivers more often about traffic accidents can paradoxically lead to an increased probability of accidents occurring, relative to leaving all drivers uninformed.
For future work, it will be interesting to situate these models in the context of network routing problems or to consider more expressive signaling policies.

\vspace{-2mm}

% Doesn't actually save that much space
% \bibliographystyle{abbrv}
\bibliographystyle{ieeetr}
\bibliography{References,library}

\end{document}